\documentclass[12pt,preprint]{aastex}

\newcommand{\myemail}{szhekov@space.bas.bg}
\newcommand{\WR}{WR~48a~}
\newcommand{\WRE}{WR~48a}
\newcommand{\kms}{~km s$^{-1}$~}

\newcommand{\XMM}{{\it XMM-Newton~}}
\newcommand{\Chandra}{{\it Chandra~}}

\shorttitle{A Chandra Observation of \WR}
\shortauthors{Zhekov et al.}

\begin{document}


\title{
A Chandra Grating Observation of the Dusty Wolf-Rayet Star \WR
}



\author{Svetozar A. Zhekov\altaffilmark{1},
Marc Gagn\'{e}\altaffilmark{2}, and 
Stephen L. Skinner\altaffilmark{3}
}

\altaffiltext{1}{Space Research and Technology Institute, Akad. G.
Bonchev str., bl.1, Sofia 1113, Bulgaria; \myemail}
\altaffiltext{2}{Department of Geology and Astronomy, West Chester
University, West Chester, PA 19383, USA;
mgagne@wcupa.edu}
\altaffiltext{3}{CASA, University of Colorado, Boulder, CO
80309-0389, USA; stephen.skinner@colorado.edu}


\begin{abstract}
We present results of a \Chandra High Energy Transmission Grating (HETG)
observation of the carbon-rich
Wolf-Rayet (WR) star \WRE. These are the first high-resolution spectra
of this object in X-rays. Blue-shifted centroids of the spectral lines
of $\sim -360$\kms and line widths of $1000 - 1500$\kms (FWHM) were deduced
from the analysis of the line profiles of strong emission lines.
The forbidden line of Si XIII is strong and not suppressed, indicating
that the rarified 10-30 MK plasma forms far from strong sources of 
far-UV emission, most likely in a wind collision zone.
Global spectral modeling showed that the X-ray spectrum of \WR
suffered higher absorption in the October 2012  \Chandra observation
compared to a previous  January 2008  \XMM observation.
The emission measure of the hot plasma in \WR
decreased by a factor $\sim 3$ over the same period of time.
The most likely physical picture that emerges from the analysis of
the available X-ray data is that of colliding stellar winds in a wide
binary system with an elliptical orbit. We propose that the unseen 
secondary star in the system is another WR star or perhaps 
a luminous blue variable.

\end{abstract}





\section{Introduction}
\WR is a carbon-rich (WC) Wolf-Rayet  star with a WC8 spectral
classification \citep{vdh_01} that was discovered in a
near-infrared survey by \citet{danks_83}. This WC star is located
inside the G305 star-forming region in the Scutum Crux arm of the
Galaxy. Its proximity (within $2'$) to the two compact infrared
clusters Danks 1 and 2 suggests that \WR likely originates
from one or the other \citep{danks_84}.
The optical extinction toward \WR is very high, A$_V = 9.2$~mag
\citep{danks_83}, and only a small part of it is due to circumstellar 
material \citep{baume_09}.
The distance to \WR is
not yet well-constrained and various studies provide a range
of $1.21 - 4$~kpc  (e.g., \citealt{vdh_01};
\citealt{baume_09}; \citealt{danks_83}).

The evolution of its infrared emission suggests that \WR is a
long-period  episodic dust-maker \citep{williams_95}. This was 
confirmed by a recent study that revealed  recurrent dust formation 
on a time scale of more than 32 years which also indicates that \WR 
is very likely a wide colliding-wind binary \citep{williams_12}.
However,  the interpretation  of the nature of \WR may not be straightforward.
Recently, \citet{hindson_12} reported the detection of a {\it thermal}
radio source (spectral index $\alpha = 0.6$, F$_{\nu} \propto \nu^{\alpha}$) 
associated with \WRE, while as a rule  wide
colliding-wind binaries are {\it non-thermal} radio sources 
\citep{do_00}.

Nevertheless, the binary nature of \WR is further supported by its X-ray
characteristics. Analysis of the \XMM spectra
of \WR showed that its X-ray emission is of thermal origin and this
is the most X-ray luminous WR star in the Galaxy detected so
far, after the black-hole candidate Cyg X-3 \citep{zhgsk_11}.
The latter is valid provided \WR is associated with the open clusters
Danks 1 and 2 and is thus located at the distance of $\sim 4$~kpc
\citep{danks_83}.

It is important to recall that all the previous pointed X-ray
observations of a small sample of presumed single WC stars have
yielded only non-detections, demonstrating that they are X-ray
faint or X-ray quiet (\citealt{os_03}; \citealt{sk_06}).
Therefore, the high X-ray luminosity of the WC star \WR is a clear
sign that it is not a single star and it is very likely that its
enhanced X-ray emission originates from the interaction region of the
winds of two massive binary components (\citealt{pril_76};
\citealt{cherep_76}).

In this paper, we report results from the first grating observation
of the dusty WR star \WRE. In Section \ref{sec:data}, we briefly
review the \Chandra HETG observation. In Section \ref{sec:overall},
we present an overview of the X-ray spectra. In Section
\ref{sec:lines}, we analyze the profiles and ratios of strong X-ray
emission lines. In Section \ref{sec:fits}, we present the results from
the global spectral fits. We discuss the results from our analysis in
Section \ref{sec:discussion} and list our conclusions in Section
\ref{sec:conclusions}.

\section{Observations and Data Reduction}
\label{sec:data}
\WR was observed with the \Chandra   HETG and ACIS-S detector
on 2012 October 12 
(\Chandra \dataset[ADS/Sa.CXO\#obs/13636]{ObsId: 13636})
with a total effective exposure of 98.6 ksec. Following
the Science Threads for Grating Spectroscopy in the 
CIAO 4.4.1\footnote{Chandra Interactive Analysis of Observations
(CIAO), http://cxc.harvard.edu/ciao/} data analysis software, the
first-order MEG/HEG and the zeroth-order HETG spectra were
extracted\footnote{We note that the associated errors for the
grating spectra are based on \citet{gehrels_86}. This is the default 
statistical error for extracting such spectra with CIAO since as a
rule the grating spectra have a small number of counts in the 
individual spectral bins.}.
The total source counts were 2083 (MEG), 1456 (HEG) and 4478 (HETG-0),
where the MEG and HEG counts are for the $+$1 and $-$1 orders combined.
The \Chandra calibration database CALDB v.4.5.0 was used to
construct the response matrices and the ancillary response files.
For the spectral analysis in this study, we made use of 
version 11.3.2 of XSPEC \citep{Arnaud96}.

\section{An Overview of the X-ray Spectra}
\label{sec:overall}
As shown in the analysis of the \XMM data \citep{zhgsk_11},
\WR is the most X-ray luminous WR star
in the Galaxy, after the black hole candidate Cyg X-3, and its
spectrum is subject to considerable X-ray absorption. 
The 5 year time interval between the \XMM and \Chandra observations
provides an opportunity to search for any changes of the X-ray emission 
that may have occurred. From such a
comparison, we can immediately conclude that the source has become
weaker and its X-ray absorption has increased, as illustrated by the
following.

\begin{figure}[hp]
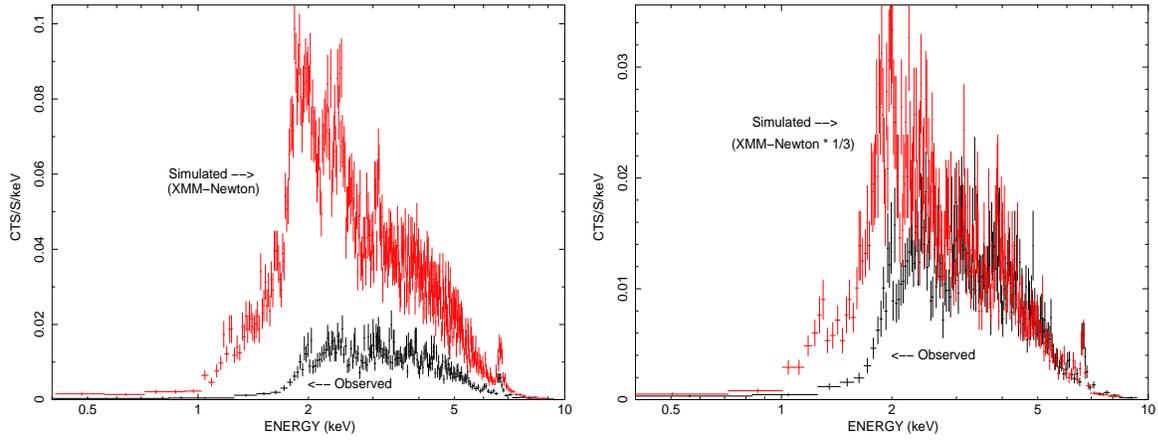

 \includegraphics[width=2.25in, height=3.in, angle=-90.]{f1a.eps}
 \includegraphics[width=2.25in, height=3.in, angle=-90.]{f1b.eps}
\caption{The \Chandra HETG-0 spectrum of \WR (black) and the
simulated spectrum in XSPEC
using the optically-thin plasma model that
perfectly fits the \XMM EPIC spectra (red). The
original spectra are shown in the left panel while in the right panel
the simulated spectrum has a reduced flux (emission measure) by a
factor of 3.
}
\label{fig:sim}
\end{figure}

Since the previous \XMM observation provided undispersed spectra 
of \WRE,  it is  natural to compare them with the undispersed HETG-0 spectrum
from the new \Chandra observation.
We used the two-shock model that perfectly fits the \XMM spectra  (see
Table~1 in \citealt{zhgsk_11}) to simulate in XSPEC the
expected HETG-0 spectrum by adopting the response matrix and the
ancillary file as derived for the \Chandra observation (see
\S~\ref{sec:data}). This procedure yields a predicted  HETG-0 spectrum
that assumes there  were no changes in the X-ray emission of 
\WR between the two observations.
Figure~\ref{fig:sim} shows two versions of the
simulated HETG-0 spectrum: one with the nominal values of the
two-shock model (absorption, temperature, emission measure) based
on the \XMM results and one
with the emission measure (intrinsic flux) decreased by a factor of 3
(absorption and temperature are kept the same). 
By comparing the left and right panels in Fig~\ref{fig:sim},
we conclude that the absorption was higher during the 2012 Chandra
observation than during the 2008 XMM-Newton observation. In addition, 
the rescaled \Chandra spectrum in Fig.\ref{fig:sim}-right shows that 
the emission measure during the Chandra observation was about a 
factor of $\sim$3 lower than during the \XMM observation.
In \S~\ref{sec:fits},
we will further
analyze and discuss this overall change in the X-ray spectrum of \WRE.





\section{Spectral Lines}
\label{sec:lines}
Due to the decrease in X-ray brightness of \WR between the two observations,
fewer counts were obtained in the \Chandra spectra than were anticipated.
As a result, we were only able to derive reliable information for a
few of the brightest lines in the grating spectra.
We re-binned the first-order MEG and HEG spectra every two bins to 
improve the photon statistics in the individual spectral 
bins\footnote{The bin sizes of the re-binned spectra are 
0.01\AA ~and 0.005\AA ~for the MEG and HEG spectra, respectively. 
The resolution element (FWHM) is 0.023\AA ~(MEG) and
0.012\AA ~(HEG): see Table 8.1 in the {\it Chandra} POG;
http://asc.harvard.edu/proposer/POG/html/index.html}.
For the Si XIII, S XV and Fe XXV He-like triplets, we fitted a sum of
three Gaussians and a constant continuum. The centers of the
triplet components were held fixed according to 
the  AtomDB data base (Atomic Data for 
Astrophysicists)\footnote{For AtomDB, see http://www.atomdb.org/}
and all components shared the same line width and line shift.
Similarly (with a sum of two Gaussians), we fitted the Si XIV and
S XVI H-like doublets but the component intensity ratios were fixed
at their atomic data values.

\begin{figure}[ht]
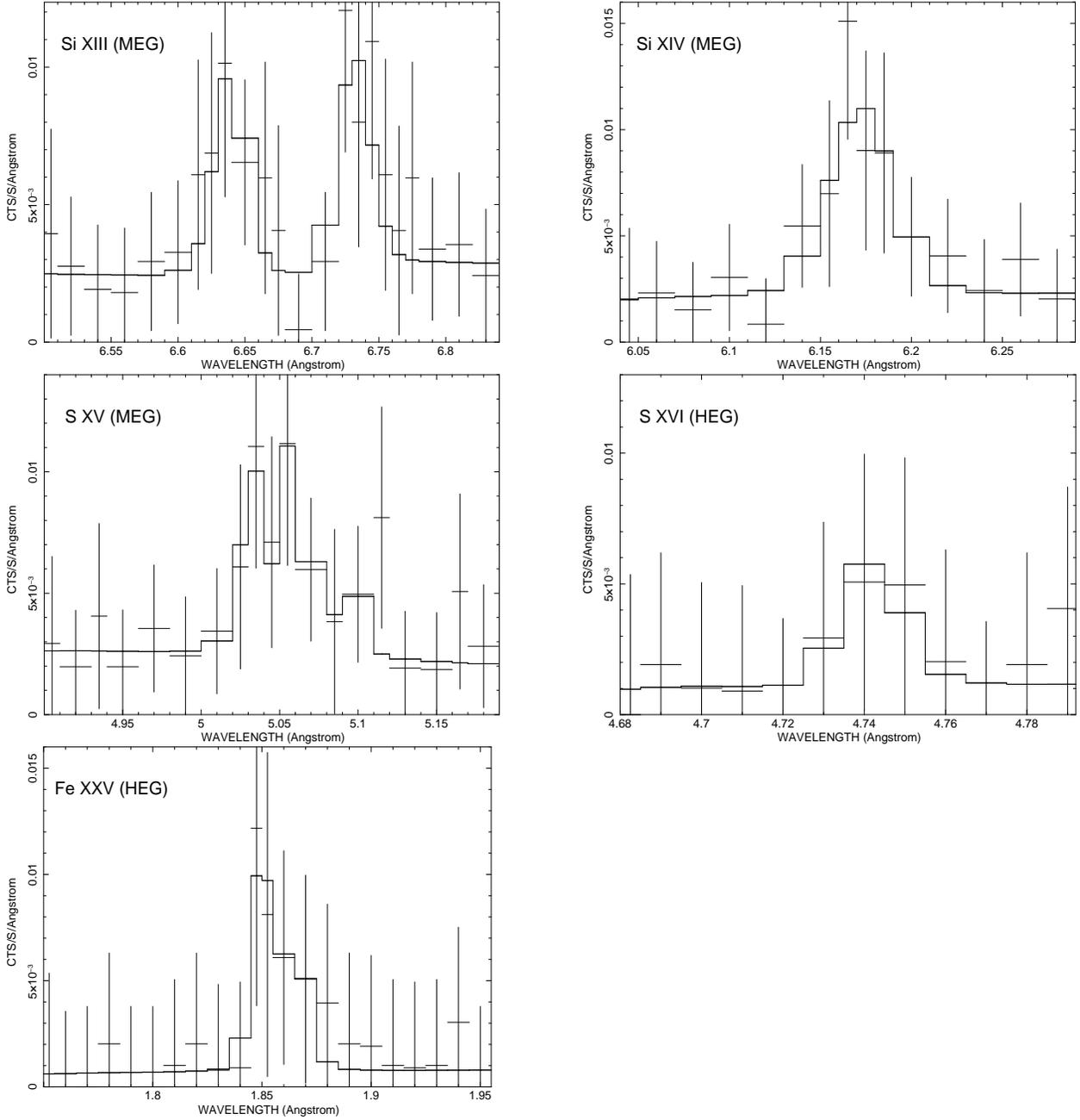

\includegraphics[width=2.25in, height=3.in, angle=-90.]{f2a.eps}
\includegraphics[width=2.25in, height=3.in, angle=-90.]{f2b.eps}
\includegraphics[width=2.25in, height=3.in, angle=-90.]{f2c.eps}
\includegraphics[width=2.25in, height=3.in, angle=-90.]{f2d.eps}
\includegraphics[width=2.25in, height=3.in, angle=-90.]{f2e.eps}
\caption{Line profile fits to some H-like doublets (Si XIV, S XVI)
and He-like triplets (Si XIII, S XV, Fe XXV) in the HEG/MEG
first-order spectra of \WRE.
The total number of counts (line$+$continuum) in all
components of a line complex (doublet or triplet)
are:
      $118\pm19$ (Si XIII),
       $81\pm15$ (Si XIV),
       $89\pm16$ (S XV),
       $16\pm11$ (S XVI),
       $28\pm12$ (Fe XXV).
For presentation purposes, the spectra were slightly re-binned with
respect to the original binning used in the fits (see
\S~\ref{sec:lines}).
}
\label{fig:profiles}
\end{figure}

\begin{figure}[ht]
\includegraphics[height=2.14in, width=3.in]{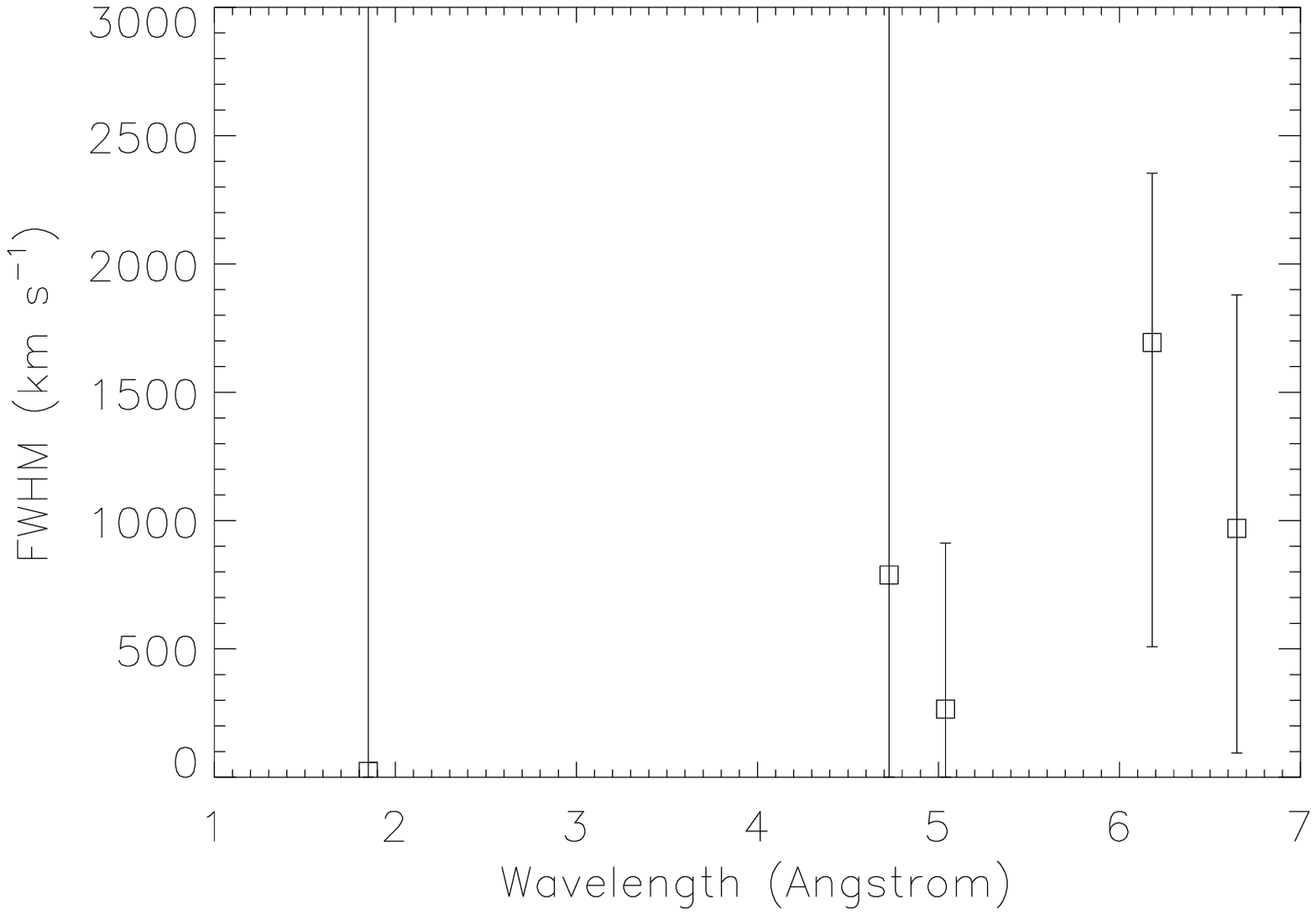}
\includegraphics[height=2.14in, width=3.in]{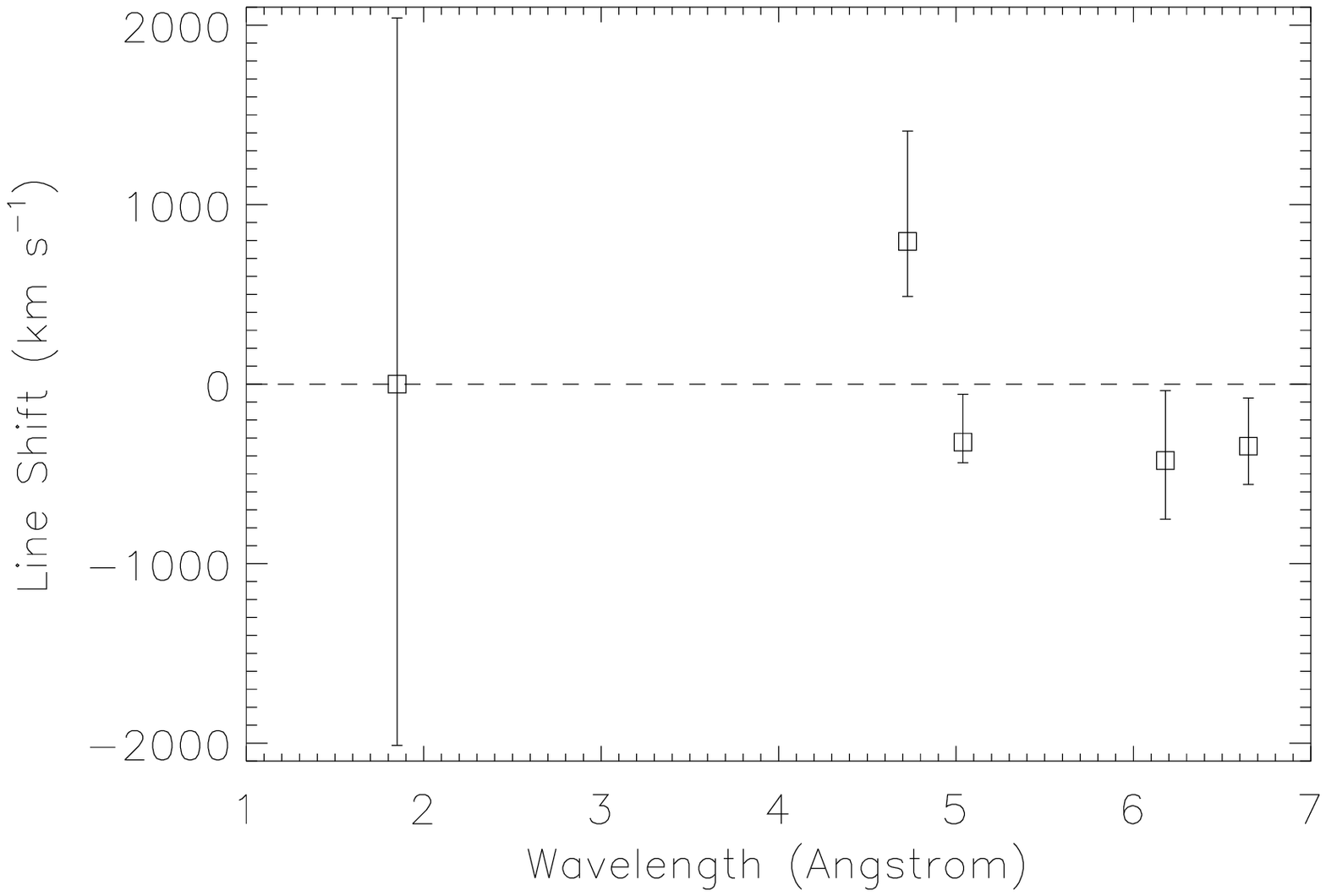}
\caption{Spectral line parameters of \WR:
Fe XXV (1.85~\AA), S XVI (4.73~\AA), S XV (5.04~\AA),
Si XIV (6.18~\AA) and Si XIII (6.65~\AA).
{\it Left} panel: full width at  half maximum (FWHM).
{\it Right} panel: the shift of the centroid of the spectral lines.
The error bars correspond to the $1\sigma$ errors from the line
profile fits (see Table~\ref{tab:lines}).
}
\label{fig:fwhm}
\end{figure}

Figures~\ref{fig:profiles},~\ref{fig:fwhm} and Table~\ref{tab:lines}
show the results from the fits to the line profiles in the X-ray grating
spectra of \WRE. Only the  Si XIII and Si XIV lines are of
acceptable quality for determining whether significant
line centroid shifts are present  (see Fig.~\ref{fig:profiles}).
These two lines  are blue-shifted in the \Chandra
observation of Oct 2012. This is further supported by results for the
the S XV line but  the data quality in that part of the
spectrum is not as good 
(e.g., near the forbidden line at $\sim 5.1$\AA). 
The other two lines which were analyzed 
show a red-shift (S XVI) or a zero-shift (Fe XXV) but this may well be
a result of their poorer photon statistics. Similarly, the Si XIII and
Si XIV line results show line broadening of about
$1000-1500$\kms, while the lines at shorter wavelengths seem to be
unresolved in the MEG and HEG spectra (e.g., their line widths are
consistent with a zero width, namely, the lower limit of the $1\sigma$
confidence interval is zero; see Table~\ref{tab:lines})
which is again likely a result of the lower data quality
(fewer counts) in these lines.

We checked all the fit results by further
improving the photon statistics in individual spectral bins. Namely,
we re-binned the spectra every four bins at the expense of some loss of 
spectral resolution: the bin size after rebinning was 0.02\AA ~(MEG) and 0.01\AA
~(HEG). The fits to these heavily re-binned spectra confirmed our basic
findings, that is blue-shifted Si XIII, Si XIV and S XV lines. As
expected from the coarser binning, the widths of all the lines were 
consistent with a zero 
width, namely, the lower limit of the $1\sigma$ confidence interval 
was zero.

We also adopted a different approach to the line profile fittting by
making use of a statistic different from the standard
$\chi^2$-statistic. 
Namely, we made use of the implementation of the Cash statistic
\citep{cash_79} in XSPEC. The fit results were very similar to those
discussed above: the fit parameters were within 5-15\% of the 
values with $\chi^2$-statistic and they were tighter constrained  
(the errors were 45-60\% of the values derived if using the standard 
$\chi^2$-statistic; see Table~\ref{tab:lines}).

Finally, to check whether the derived results from the line profile fits
 are sensitive to the `local'
level of continuum, we performed a `global-continuum' fit. Namely, we
fitted {\it simultaneously} all the lines discussed above by
introducing a common continuum for all the lines, represented by 
bremsstrahlung emission
with a plasma temperature kT = 3 keV. (Fits with a variable
plasma temperature were equally successful.) The line profiles were
modeled  as above, that is by a sum of two or three Gaussians for
the doublet and triplet lines, respectively. This `global-continuum'
fit provided line parameters that were identical  to the ones
given in Table~\ref{tab:lines}. These consistency checks  provide additional
confidence in the derived spectral line parameters despite their
relatively large uncertainties.






\section{Global Spectral Fits}
\label{sec:fits}
The infrared variability of \WR suggests that it is a long-period dust
maker, thus, it is likely a long-period WR$+$O binary system 
(\citealt{williams_95}; \citealt{williams_12}).
The \XMM observation of its X-ray
emission provided  additional support for its suspected  binary nature  
by discovering that \WR is a
luminous X-ray source with high plasma temperature \citep{zhgsk_11}.
All these characteristics indicate that colliding
stellar winds (CSW) may play an important role in the physics of this
object. We recall that as a result of CSWs a two-shock structure
forms between the two massive stars in the binary system. This structure is a
source of X-ray emission (Prilutskii \& Usov 1976; Cherepashchuk 1976)
and it is believed to give rise to the variable infrared  emission
(dust emission; e.g. \citealt{williams_90}). We will thus explore
the CSW picture in the analysis of the \Chandra spectra of \WRE. 
The best way of doing this is to confront the results from
hydrodynamic modeling of CSWs in \WR with the observations. However,
basic physical parameters needed to construct a detailed model such
as the stellar wind parameters and the binary separation are not available.
We will therefore adopt a simplified approach based on discrete-temperature
plasma models.
We emphasize that  discrete-temperature models are a
simplified representation of the temperature-stratified CSW region 
(see Section 5.2 in \citealt{zh_07} for discussion of CSW models 
versus discrete-temperature models).
In this as well is in our previous study of the X-ray emission from
\WR \citep{zhgsk_11}, we adopted the discrete-temperature model 
$vpshock$ in XSPEC. It is a physical model of the X-ray emission
behind a strong plane-parallel shock that takes into account the
effects of non-equilibrium ionization (for details of the model see
\citealt{borkowski_01}).

\begin{figure}[ht]
\includegraphics[width=5.6in, height=4.in]{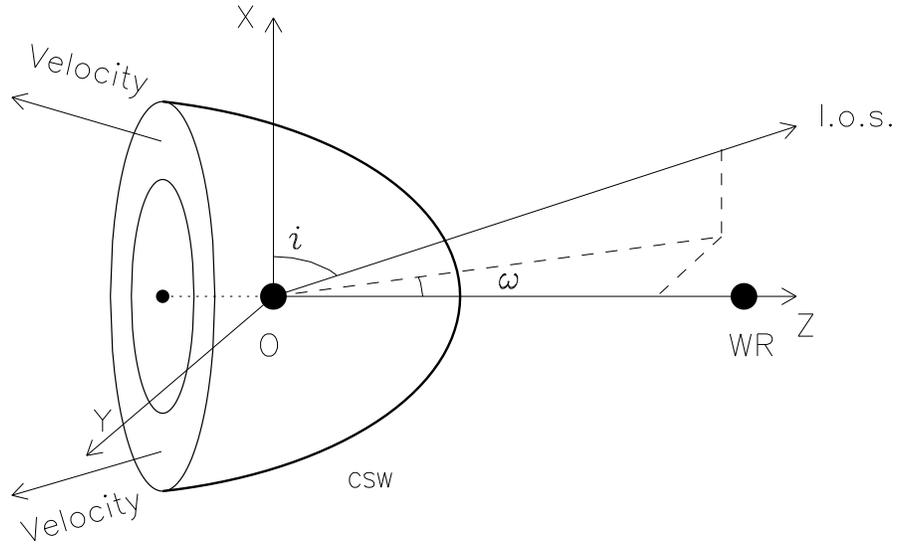}
\caption{
A schematic diagram of colliding stellar winds in
a massive WR$+$O binary system.
The wind interaction `cone'
is denoted by CSW
(the axis Z is its axis of symmetry;
the axis X is perpendicular to the orbital plane;
the axis Y completes the right-handed coordinate system).
The line-of-sight towards observer is denoted by l.o.s. and
the two related angles, $i$ (orbital inclination) and $\omega$
(azimuthal angle) are marked as well. The arrows labeled
`Velocity' indicate the general direction of the bulk gas velocity in
the
interaction region.
}
\label{fig:cartoon_DD}
\end{figure}

As demonstrated in \S~\ref{sec:overall}, apparent changes have
occurred in the X-ray emission of \WR between  January 2008 
October 2012.  Namely,
the X-ray emission became weaker and more absorbed 
(Fig.~\ref{fig:sim}). Such changes are possible in the
standard CSW picture even for steady-state  stellar winds 
(i.e. mass-loss rates and wind velocities are constant in time).
For example, if the orbital inclination is considerably higher than zero
degrees then  higher X-ray absorption should be detected for orbital
phases when the star with the more massive wind in the binary is
`in front' 
(e.g., at azimuthal angles around $\omega \approx 0\arcdeg$; 
see Fig.~\ref{fig:cartoon_DD}), 
In a WR$+$O binary this occurs when the WR star is located 
between the observer and the CSW region. This change (increase)
in absorption will occur when the WR star is `in front' regardless
of whether the orbit is circular or elliptical.
On the other hand, the intrinsic X-ray luminosity of CSWs will
change during the orbit only in the case where the orbit is elliptical.
We recall 
that there exists a scaling law for the CSW X-ray luminosity with the
mass-loss rate ($\dot{M}$), wind velocity ($v$) and binary separation
($D$): $L_X \propto \dot{M}^2 v^{-3} D^{-1}$ (\citealt{luo_90};
\citealt{mzh_93}). We thus see that even for constant
mass-loss rate and wind velocity the intrinsic CSW luminosity will
vary due to the change of the binary separation if the orbit is
elliptical.  $L_X$ will reach its maximum near periastron 
(at the minimum value of $D$) and will have its
minimum at apastron (the maximum value of $D$).
We will next explore the CSW picture in the analysis of the \Chandra
spectra of \WR by assuming that (most of) its X-ray emission
originates in CSWs.

We note that all of the global spectral models considered in this
study have the following basic characteristics. 
Since no appreciable excess in the X-ray absorption above that
expected from the ISM was found in the analysis of the \XMM
spectra of \WRE, the column density of the
interstellar X-ray absorption was kept fixed to
N$_{H, ISM} = 2.30\times10^{22}$~cm$^{-2}$ (see Table 1 in
\citealt{zhgsk_11}). The additional X-ray absorption that is
observed in the \Chandra spectra (see \S~\ref{sec:overall} and
Fig.~\ref{fig:sim}) is attributed to the wind absorption in the more
massive WR wind in the presumed WR$+$O binary. 
The {\it vphabs} model in XSPEC was used for this additional absorption
component and we enforced
its abundances to be
the same as that of the emission component(s).
The abundances of the emission component(s) are with respect to the
typical WC abundances \citep{vdh_86} and they were kept fixed to their 
values derived in Zhekov, Gagn\'{e}, \& Skinner (2011; see the 
2T-shock model in Table 1 therein)\footnote{
Abundances in XSPEC are set by the $abund$ command which reads a
default file
that nominally contains the fractional abundance by number of each
element
relative to hydrogen (e.g. for solar abundances H = 1.0, He = 0.097,
etc.).
To simulate  nonsolar abundances for He-rich but H-depleted WC stars
the default abundance file is modified such that He is assigned a very
high abundance relative to H (H = 1.0, He = 50. in our simulations)
and the remaining elements are then scaled
to their appropriate number fractions  relative to He based on
the values  given in 
van der Hucht et al. (1986).
}
These values were (expressed as scaling factors to be applied to
the typical WC abundances):
He $= 1$, C $= 1$, N $= 1$,
Ne $= 0.11$, Mg $= 0.13$, Si $= 0.64$, S $= 1.78$, Ar $= 2.78$,
Ca $= 1.97$, Fe $= 1.31$.
In our analysis, the total (summed) first-order MEG and HEG spectra and the zeroth-order
HETG spectra were fitted simultaneously. To improve the photon
statistics, the spectra were re-binned to have a minimum of 20 counts
per bin. Based on the results from the
line profile analysis (\S~\ref{sec:lines}), a typical line broadening
of
FWHM $= 1000$\kms and a typical line shift of $- 365$\kms were
assumed.

Let us now attribute the entire X-ray emission from \WR to the CSWs in
a presumed WR$+$O binary. We recall that if the binary orbit is circular, no
changes of the intrinsic X-ray parameters are expected
with the orbital phase  while the X-ray
luminosity may vary if the orbit is elliptical. We denote these two
cases
CSW$_{circ}$  and CSW$_{ell}$, respectively.

{\it CSW$_{circ}$}.
In this case, we made use of the two-shock model that successfully
fitted the \XMM spectra of \WR (see the 2T-shock model in Table 1
in \citealt{zhgsk_11}; the model uses the
{\it vpshock} model in XSPEC in conjunction with the XSPEC
command $xset~neivers~ 2.0$ to select version 2.0 of the
non-equilibrium ionization collisional plasma model). 
All the model parameters were
kept fixed to their values as derived in that analysis. Since 
excess
X-ray absorption is seen in the \Chandra data (see
\S~\ref{sec:overall} and  Fig.~\ref{fig:sim}), the two shock
components suffer additional {\it wind} absorption as described above.
We first assumed that both emission components have equal
wind absorptions. This resulted in a very poor-quality spectral fit 
(reduced $\chi^2$  $ \approx 9$). If different X-ray
absorptions for each component were allowed, the quality of the fit
improved but was still statistically unacceptable (reduced
$\chi^2 = 3.2$; see Table~\ref{tab:fits} for details).
We can thus conclude that the physical picture of CSW in a WR$+$O
binary that has a {\em circular}  orbit is not supported by the X-ray
observations of \WRE. Namely, such a picture cannot explain the
observed changes in the X-ray emission from this object.

\begin{figure}[ht]
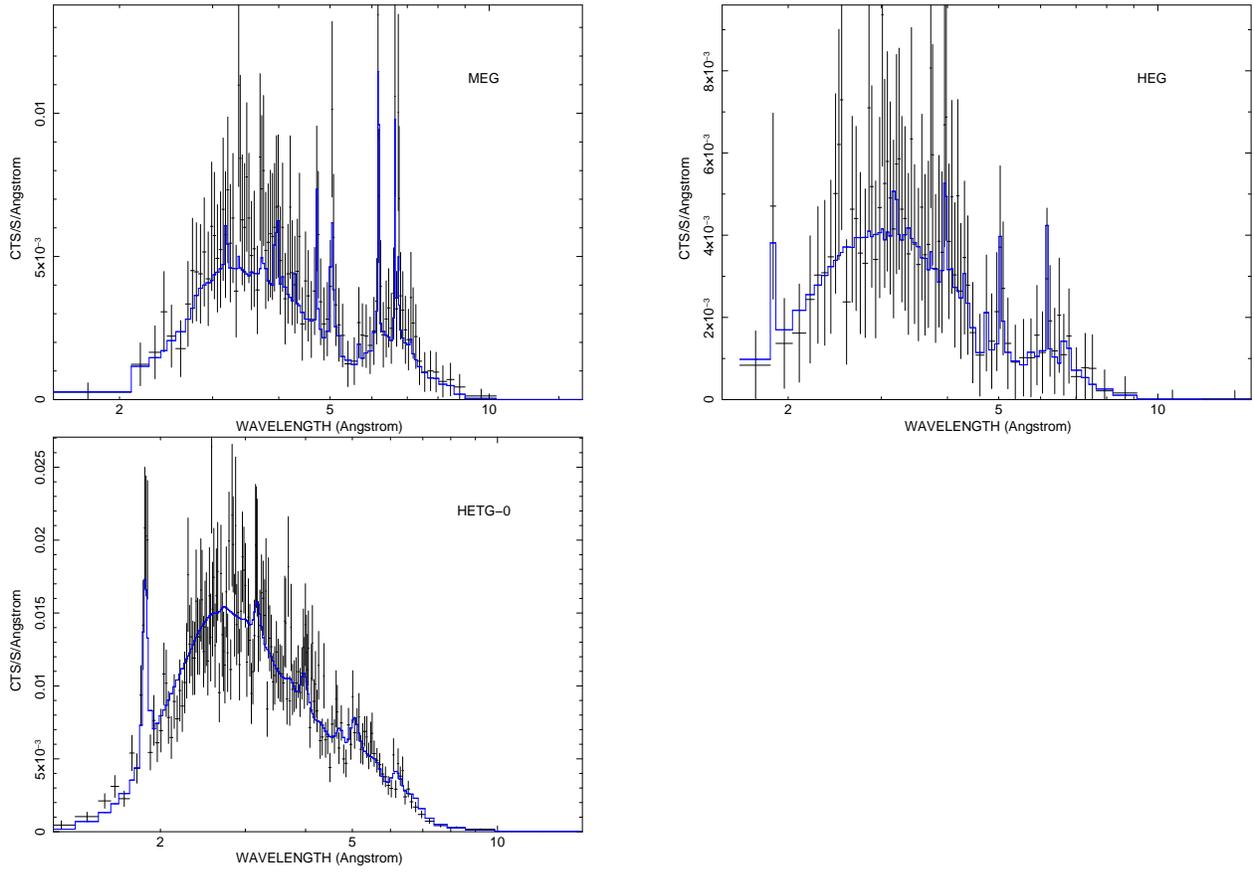

 \includegraphics[width=2.25in, height=3.in, angle=-90.]{f5a.eps}
 \includegraphics[width=2.25in, height=3.in, angle=-90.]{f5b.eps}
 \includegraphics[width=2.25in, height=3.in, angle=-90.]{f5c.eps}
\caption{The HETG background-subtracted spectra of \WR and
the two-component model fit (see model CSW$_{ell}$ in
Table~\ref{tab:fits}). The spectra were re-binned to have a minimum
of 20 counts per bin.
}
\label{fig:fits}
\end{figure}

{\it CSW$_{ell}$}.
In this case, we made use of the same model as above (CSW$_{circ}$)
but we also allowed the shock emission measure to vary. This means
that we assumed that the shape of the intrinsic X-ray emission does
not change with the orbital phase and only the amount of the X-ray
emitting plasma does. In other words, although the distance between
the stars in the binary varies it does not become very small, so, the
stellar winds still  have enough space to reach their terminal velocities
before they collide. We believe that this is a reasonable assumption
for a wide binary system having a long period orbit as is likely the case
with \WR (its orbital period is at least of 32 years; e.g.,
\citealt{williams_12}). In the fitting procedure, we kept all the 
shock plasma parameters fixed to their values in the original 
two-shock model (see the 2T-shock model in Table 1 in
\citealt{zhgsk_11}). We allowed the total emission measure to vary 
but we kept the ratio of the emission measures of the shock components 
fixed to its value as derived in the analysis of the \XMM spectra.
By fixing this ratio, the shape of the X-ray spectrum is not allowed
to vary.
Both models with a common (equal) or with a different
X-ray absorption for each shock component gave very good fits to the
observed \Chandra HETG spectra. Figure~\ref{fig:fits} and
Table~\ref{tab:fits} present the results for the case with a separate
X-ray absorption for each emission component (i.e. for each shock). 
The model fit with equal wind absorptions had $\chi^2$/dof $= 251/349$ and
it gave identical plasma parameters with the ones presented
in Table~\ref{tab:fits}. The common wind absorption was practically a
mean of the values given in that table, namely
N$_{He, wind} = 0.16^{+0.01}_{-0.01}\times 10^{21}$ cm$^{-2}$.
The basic result from the CSW$_{ell}$ case is that in October 2012 the
intrinsic X-ray emission (the amount/emission measure of hot plasma
as well) of \WR was by a factor $\sim 3$ lower than its value in January
2008 and the level of the X-ray attenuation  increased at the same
time, as anticipated in \S~\ref{sec:overall}.

In the framework of the standard CSW picture in a binary with
an elliptical orbit, the lower X-ray luminosity should be observed when
the
 binary separation is larger. This means that the orbital orientation
of \WR is such that in October 2012 the binary separation was larger than
it was in January 2008. To have an increased value of the X-ray
attenuation, the star with the more massive wind should be nearly `in
front' in October 2012 
(i.e. azimuthal angles around $\omega \approx 0\arcdeg$; 
see Fig.~\ref{fig:cartoon_DD}), 
so then its counterpart was likely `in front' in January 2008
(i.e. azimuthal angles around $\omega \approx 180\arcdeg$;          
see Fig.~\ref{fig:cartoon_DD}). This seems to be the case
since the X-ray absorption (neutral hydrogen column density)
showed no appreciable excess above the interstellar value in January 2008
\citep{zhgsk_11}. However, if the WR star is `in
front' the plasma in the interaction region will be outflowing
(see Fig.~\ref{fig:cartoon_DD}), that
is it is moving away from the observer. Since the WR  wind is expected to be dominant in
the system,  the CSW `cone' will have its apex pointing towards
the WR star (and the observer as well). But instead  we found
blue-shifted spectral lines in the X-ray spectrum of \WR in October 2012
(see \S~\ref{sec:lines}) which indicates that the X-ray emitting
plasma was moving towards the observer. This suggests that the star with
the weaker  wind was `in front' at the time of the
\Chandra observation in October 2012
(i.e. azimuthal angles around $\omega \approx 180\arcdeg$;
see Fig.~\ref{fig:cartoon_DD}). 
We will return to this apparent discrepancy
in \S~\ref{sec:discussion}. 

\section{Discussion}
\label{sec:discussion}
The most important results from the analysis of the \Chandra HETG
spectra of \WR \, are the following: (i) blue-shifted and broadened
(spectrally resolved) emission lines are
detected; (ii) in October 2012, the  emission measure  of the hot gas
is about a factor of $\sim 3$ lower than it was in January 2008;
(iii) the X-ray absorption has increased considerably over the same
period of time.

As discussed above (\S~\ref{sec:fits}), the CSW picture was adopted in
our analysis of the entire X-ray emission from \WRE. It was done in a
simplified manner by making use of a two-shock model. We recall that
the CSW region is temperature-stratified and the two-temperature
plasma (two-shock model) is just a simplified representation of the
temperature stratification of its hot plasma weighted by its emission
measure. Thus, we believe that the results derived from the two-shock
modeling (the last two mentioned above) are definitely valid in the
CSW picture. To double check these two results, we also fitted the
HETG spectra with a one-shock model. The results from this model
fitting also confirmed the decrease of emission measure by the same
factor ($\sim 3$) and the increased (wind) absorption (see 1T-shock
model in Table~\ref{tab:fits}). It is worth noting that the
success of the single-temperature shock model in this case 
is due to the fact that the X-ray spectrum of \WR is heavily absorbed.
However, we have given preference to the two-shock model in our analysis
(\S~\ref{sec:fits}) for consistency with the previous
study of the X-ray emission of \WR in which a two-shock model
was required to match the observed spectra successfully
\citep{zhgsk_11}.

As already mentioned, each of the three most important results from
this study can be  explained in the framework of the CSW picture.
Namely, the blue-shifted emission lines indicate that the star with
the less powerful wind was globally `in front' of the CSW region at
the time of the \Chandra observation. On the other hand, the decrease
of the emission measure over a period of almost
5 years  is a sign of an elliptical orbit
in the presumed WR$+$O binary system. Finally, the increased X-ray
absorption in October 2012  compared to that in January 2008 indicates that we
observed the X-ray emission from the CSW region through a massive
stellar wind in the former and likely through the hot gas of the CSW
region (or at an orbital phase far from conjunction, thus, no wind
absorption) in the latter.
But, can we get simultaneously these three observational facts to be
consistent with the CSW picture? We believe it is possible in the
following, although quite speculative, way.

Suppose the binary system in \WR consists of two massive stars: a WR
star (of the WC subtype) being the primary in the system and a
secondary star which  is not an O star but instead  a WR or a
LBV (luminous blue variable) star. (The latter LBV case might
be an object similar to the SMC star 
HD5980;  see \citealt{gloria_10} and references therein.) 
Suppose the stellar wind of the WC primary is more powerful
than that of the secondary. In the case of an elliptical orbit, the
decrease of the emission measure of the CSW region between January 2008
and October 2012 is then explained by the increase of the binary
separation over that period of time. If the secondary star was
`in front' 
(e.g., azimuthal angles around $\omega \approx 180\arcdeg$;
see Fig.~\ref{fig:cartoon_DD})
at the time of the \Chandra observation (October
2012), then this would explain the detection of blue-shifted lines in
the spectrum of \WRE. On the other hand, if the secondary is a WR
or LBV star the high X-ray {\it wind} absorption is also expected at
this orbital phase provided the orbital inclination is high enough.
In fact, if the latter is close to 90 degrees and the orbital phase
at which  the secondary is `in front' has not yet passed, the X-ray absorption
will keep increasing. Alternatively, if that orbital phase has already
passed the X-ray absorption is already beyond its maximum and it will
be decreasing in the future. This would be  a secondary maximum
for the X-ray absorption since its primary maximum will occur when the
primary star (WC) is `in front'
(e.g., azimuthal angles around $\omega \approx 0\arcdeg$;
see Fig.~\ref{fig:cartoon_DD}). 
And to explain the low
(negligible) {\it wind} absorption in the January 2008 \XMM observation, we
should assume that at that time we observed the CSW region through its
hot gas. This means that the line-of-sight intercepted the CSW `cone'
itself (bounded by the two shock surfaces) which eliminates the wind
absorption. If this were the case, we may expect that after the
phases with increased X-ray absorption have passed (see above) we will witness
another phase with low wind absorption
which will allow us to directly estimate the
opening angle of the CSW `cone'.
It is worth noting that the chemical composition of the absorbing wind
cannot be constrained from the X-ray data alone. We recall
that in our analysis the wind absorption component and the emission
component share the same abundances. However, successful fits to the
HETG spectra were possible even if we assumed that the wind absorption
component had a chemical composition typical for  interstellar
matter. Therefore, our suggestion that the secondary star in the
binary might be a WR or a LBV star is not in conflict with the data.
Nevertheless, it is worth noting that deep high-resolution optical 
observations are essential to validate this suggestion.

Finally, we note that the observed changes in the X-ray emission of
\WR between January 2008 and October 2012 might well be due to some temporary
state of the binary system, e.g., they could be a consequence of 
time-variable  stellar wind parameters.
Asymmetric stellar winds, having an appreciable contrast of the wind
parameters between the stellar pole and equator, could also contribute
to those changes.
Also, \WR could be a system of higher hierarchy than just a wide
binary system with a period of $\geq 32$ years as proposed by
\citet{williams_12}.
But for interpreting the data in hand,
such explanations seem more
speculative to us than the one described above. We note
that as in the \XMM data \citep{zhgsk_11} we found
no short-term variability in the
\Chandra data of \WRE.  On a timescale less than 100 ksec and time bins
between 100 and 2000 s, the X-ray light curve is statistically 
consistent with a constant flux.
On the other hand, even if the physical picture behind the X-ray
emission from \WR were more complicated than just CSWs, the latter
is likely  one of its important ingredients. The strong 
forbidden line in the He-like triplet of Si XIII
(Table~\ref{tab:lines}) is a solid indicator that this line complex
forms in a rarefied hot plasma far from strong UV sources which
clearly  points to a CSW region in a wide massive binary system.
It is worth noting that the other   He-like triplet detected in the \WR
spectra, namely S XV, shows enhanced emission from the intercombination
line which may indicate a different origin of the very hot plasma
where the S XV  line complex forms. However, as already mentioned 
(\S~\ref{sec:lines}) the data in that part of the spectrum are of
lower quality and a deeper observation would be needed to 
validate this result.

\section{Conclusions}
\label{sec:conclusions}
In this work, we presented an analysis of the first grating X-ray
spectra of the dusty WR star \WR obtained with the \Chandra HETG.
The basic results and conclusions are as follows.

\begin{enumerate}
\item
From analysis of the line profiles of strong emission lines, a typical
line width (FWHM) of $1000 - 1500$\kms was deduced and blue-shifted
line centroids of $\sim -360$\kms. A strong (not suppressed) 
forbidden line in the He-like triplet of Si XIII was detected which 
indicates that this line forms a rarefied hot plasma far from strong 
sources of UV emission.

\item
Global spectral modeling showed that the X-ray spectrum of \WR
suffered higher absorption (likely of {\it wind} origin) in October 2012 
(the \Chandra observation) compared to January 2008
(the \XMM observation). The emission measure
of the hot plasma in \WR decreased by a factor $\sim 3$ over the same
period of time.

\item
No X-ray variability on a timescale of less than 100 ksec was detected.
This result is similar to what was found from the analysis of the
previous X-ray observation of \WR with \XMM \citep{zhgsk_11}.

\item
The most likely physical picture that emerges from the analysis of
the available X-ray data (\Chandra and \XMM) is the following. The
high X-ray luminosity of the carbon-rich WR star \WR is due to
colliding stellar winds in a wide binary system with elliptical
orbit. The observed changes of the characteristics (emission 
measure, absorption) of the X-ray emission of \WR and the
blue-shift of the line centroids of the strong lines can find their
place in the CSW picture, provided the secondary star in the binary
system (the companion star of the primary WC object) is not an O star
but instead a WR star or a LBV star.

\item
More X-ray observations with high spectral resolution
are needed to describe in detail the X-ray characteristics of \WRE.
Due to the high X-ray absorption of this object (most of the
X-ray emission from \WR emerges at energies above 2 keV), 
\Chandra
HETG data are essential for us to obtain a deeper
understanding of the physical picture of CSWs in the fascinating
dusty WR star \WRE.

\end{enumerate}




\acknowledgments
This work was supported by NASA through {\it Chandra} Award GO2-13017X
to  West Chester University, West Chester, Pennsylvania.
SAZ acknowledges financial support from Bulgarian National Science
Fund grant DO-02-85.
The authors thank an anonymous referee for 
helpful comments and suggestions.



{\it Facilities:} \facility{{\it Chandra} (HETG)}.

\clearpage



%

\clearpage

\clearpage

\clearpage

\clearpage

\clearpage

\begin{deluxetable}{lcrrrc}
\tablecaption{Line Parameters
\label{tab:lines}}
\tablewidth{0pt}
\tablehead{
\colhead{Line} & \colhead{$\lambda_{lab}^{a}$} & \colhead{FWHM$^{b}$} & 
\colhead{Line Shift$^{c}$} & \colhead{Flux$^{d}$} & \colhead{Ratio}\\
\colhead{} & \colhead{(\AA)} &  \colhead{(\kms)} &  
\colhead{(\kms)} & \colhead{}  &  \colhead{(ATOMDB)}
}
\startdata
Fe XXV K$_{\alpha}$ & 1.850 &
23$^{+6001}_{-23}$ & 0$^{+2040}_{-2013}$ & 12.26$^{+5.59}_{-6.40}$ \\
\,\,\,(i/r)$^{e}$  &  &  &  & 0.43$^{+....}_{-....}$ & 0.38 \\
\,\,\,(f/r)$^{e}$  &  &  &  & 0.48$^{+....}_{-....}$ & 0.30 \\
  &  & 
(1460$^{+838}_{-864}$) & (605$^{+309}_{-332}$) & 
                       (12.82$^{+3.46}_{-2.59}$) \\
  &  &  &  & (0.00$^{+0.44}_{-0.00}$) & 0.38 \\
  &  &  &  & (0.47$^{+0.43}_{-0.21}$) & 0.30 \\
  &  &  &  &  &  \\
S XVI L$_{\alpha}$ & 4.727 &
788$^{+2914}_{-788}$ & 795$^{+614}_{-307}$ & 5.32$^{+6.12}_{-4.79}$ \\
  &  &  
(794$^{+575}_{-539}$) & (773$^{+223}_{-215}$) &
                      (5.24$^{+2.66}_{-1.81}$) \\
  &  &  &  &  &  \\
S XV K$_{\alpha}$ & 5.039 &
265$^{+646}_{-265}$ & -323$^{+267}_{-115}$ & 12.10$^{+4.81}_{-4.11}$
\\
\,\,\,(i/r) &  &  &  & 1.27$^{+2.11}_{-0.69}$  &  0.23 \\
\,\,\,(f/r)  &  &  &  & 0.49$^{+1.37}_{-0.49}$ &  0.44 \\
  &  &
(186$^{+592}_{-186}$) & (-356$^{+13}_{-137}$) &
                        (11.83$^{+2.89}_{-2.41}$) \\
  &  &  &  & (1.63$^{+1.46}_{-0.61}$)  & 0.23  \\
  &  &  &  & (0.74$^{+0.76}_{-0.34}$)  & 0.44  \\
  &  &  &  &  &  \\
Si XIV L$_{\alpha}$ & 6.180 &
1694$^{+659}_{-1186}$  & -426$^{+390}_{-327}$ &
3.99$^{+2.01}_{-1.48}$\\
  &  &
(1571$^{+963}_{-684}$)  & (-461$^{+215}_{-222}$) &
                          (4.30$^{+1.12}_{-1.10}$)  \\
  &  &  &  &  &  \\
Si XIII K$_{\alpha}$ & 6.648 &
970$^{+909}_{-875}$ & -346$^{+268}_{-213}$ & 4.71$^{+2.07}_{-1.76}$ \\
\,\,\,(i/r)  &  &  &  & 0.00$^{+0.22}_{-0.00}$ &  0.20 \\
\,\,\,(f/r)  &  &  &  & 1.03$^{+1.01}_{-0.52}$ &  0.52 \\
  &  &
(869$^{+402}_{-372}$) & (-307$^{+105}_{-115}$) & 
                        (5.01$^{+1.12}_{-0.96}$) \\
  &  &  &  & (0.00$^{+0.11}_{-0.00}$) &  0.20 \\
  &  &  &  & (1.07$^{+0.52}_{-0.31}$) &  0.52 \\
\enddata
\tablecomments{
Results from the fits to the line profiles in the first-order MEG
(Si XIII, Si XIV, S XV) and
HEG (S XVI, Fe XXV)
spectra with the associated $1\sigma$ errors 
($1\sigma$ error is equivalent to a change in the best-fit statistic
value by 1).
For the He-like triplets,
the flux ratios of the intercombination to the resonance line (i/r) and of
the
forbidden to the resonance line (f/r) are given as well.
The errors on the spectral parameters are those following the
\cite{gehrels_86} recommendation for the cases with a small number of
counts in individual data bins (see \S \ref{sec:data}). When we enforced
the Gaussian errors on the data (error $= \sqrt{N}$, $N$ is
the number of counts in a data bin), the results from the fits to the
line profiles remained within 10-15\% of the values given in this Table
and the errors from the fits were 40-60\% of the values listed here.
For comparison in parentheses, given are the results from the fits
based on the Cash statistic \citep{cash_79}.
\\
$^{a}$ The laboratory wavelength of the main component.\\
$^{b}$ The line width (FWHM).\\
$^{c}$ The shift of the spectral line centroid.\\
$^{d}$ The observed total multiplet flux in units of $10^{-6}$
photons cm$^{-2}$ s$^{-1}$.\\
$^{e}$ Due to the poor photon statistics, these line ratios are not
constrained.
}

\end{deluxetable}

\clearpage

\begin{deluxetable}{llll}
\tablecaption{Global Spectral Model Results
\label{tab:fits}}
\tablewidth{0pt}
\tablehead{
\colhead{Parameter} & \colhead{CSW$_{circ}$}  & \colhead{CSW$_{ell}$} &
\colhead{1T
shock}
}
\startdata
$\chi^2$/dof  & 1107/349 & 249/348 & 249/347 \\
N$_{H, ISM}$ (10$^{22}$ cm$^{-2}$)  &
          2.30 & 2.30 &  2.30 \\
N$_{He, wind,1}$ (10$^{21}$ cm$^{-2}$)  &
          0.12$^{+0.01}_{-0.01}$ & 0.35$^{+0.06}_{-0.14}$ &
          0.13$^{+0.02}_{-0.01}$  \\
N$_{He, wind,2}$ (10$^{21}$ cm$^{-2}$)  &
          2.65$^{+0.11}_{-0.07}$ & 0.12$^{+0.16}_{-0.07}$ & \\
kT$_1$ (keV) & 1.05 & 1.05 & 2.41$^{+0.09}_{-0.14}$  \\
kT$_2$ (keV) & 2.82 & 2.82 & \\
EM$_1$ ($10^{54}$~cm$^{-3}$) &  2.42 &  0.88$^{+0.02}_{-0.02}$ &
                                        2.40$^{+0.17}_{-0.04}$  \\
EM$_2$ ($10^{54}$~cm$^{-3}$) &  5.34  & 1.94 &  \\
$\tau_1$ ($10^{11}$ cm$^{-3}$ s)  & 2.42  &  2.42 &
                                            15.9$^{+6.70}_{-4.10}$  \\
$\tau_2$ ($10^{11}$ cm$^{-3}$ s)  & 8.09  & 8.09 &  \\
F$_{X}$ ($10^{-11}$ ergs cm$^{-2}$ s$^{-1}$)  &
           0.30 (17.7) & 0.28 (6.44) &  0.27 (2.09) \\
F$_{X,hot}$ ($10^{-11}$ ergs cm$^{-2}$ s$^{-1}$)  &
           0.20 (6.98) & 0.26 (2.53) &  \\
\enddata
\tablecomments{
Results from  simultaneous fits to the \Chandra HETG
spectra (including both 0-order and 1st-order spectra) of \WR
(the plasma model used in the fits is the $vpshock$ model in XSPEC;
see \S~\ref{sec:fits} for details).
The abundances were with respect to the
typical WC abundances \citep{vdh_86} and  kept fixed to their
values derived in \citet{zhgsk_11}.
Tabulated quantities are the neutral hydrogen absorption column
density (N$_{H, ISM}$), the neutral helium absorption column density
(wind absorption; N$_{He, wind}$), plasma temperature (kT),
emission measure ($\mbox{EM} = \int n_e n_{He} dV $)
for a reference distance of d~$=1$~ kpc ($\mbox{EM}\propto d^2$),
shock ionization age
($\tau = n_e t$), the absorbed X-ray flux (F$_X$) in the
0.5 - 10 keV range followed in parentheses by the unabsorbed value
(F$_{X,hot}$~ denotes the higher-temperature component).
Errors are the $1\sigma$ values from the fits.
}

\end{deluxetable}




\end{document}